
\input harvmac
\input psfig
\noblackbox
%
%
%

\def\Title#1#2{\ifx\answ\bigans \nopagenumbers
\abstractfont\hsize=\hstitle\rightline{#1}%
\vskip .5in\centerline{\titlefont #2}\abstractfont\vskip .5in\pageno=0
\else \rightline{#1}
\vskip .8in\centerline{\titlefont #2}
\vskip .5in\pageno=1\fi}
\ifx\answ\bigans

\else

 \font\absi=cmmi10 scaled\magstep1
\font\absis=cmmi7 scaled\magstep1 \font\absiss=cmmi5 scaled\magstep1
\font\abssy=cmsy10 scaled\magstep1 \font\abssys=cmsy7 scaled\magstep1
\font\abssyss=cmsy5 scaled\magstep1 
\skewchar\absi='177 \skewchar\absis='177 \skewchar\absiss='177
\skewchar\abssy='60 \skewchar\abssys='60 \skewchar\abssyss='60
\fi
%
%
\font\ticp=cmcsc10

\def\ajou#1&#2(#3){\ \sl#1\bf#2\rm(19#3)}

\def\frac#1#2{{#1 \over #2}}
\def\vev#1{\langle#1\rangle}
%
\lref\bholeref{S. Elitzur, A. Forge, and E. Rabinovici, ``Some Global Aspects
of String Compactifications,'' preprint RI-131 (1990) \semi
K. Bardacki, M. Crescimanno, and E. Rabinovici, ``Parafermions
from Coset Models,'' LBL preprint (1990) \semi
M. Rocek, K. Schoutens, and A. Sevrin, ``Off-Shell WZW Moels in
Extended Superspace,'' IASSNS-HEP-91/14 \semi
G. Mandal, A. Sengupta, and S. Wadia, ``Classical Solutions of
2-Dimensional String Theory, '' IASSNS-HEP-91/10.}
\lref\wittenlec{E. Witten, ``On Black Holes in String Theory,'' Lecture
at Strings '91, Stonybrook, 1991.}
\lref\Waldcmp{R.M. Wald, ``On Particle Creation by Black Holes,''
\ajou Commun. Math. Phys. &45 (75) 9.}
\lref\unpred{S.W. Hawking, ``Breakdown of Predictability
in Gravitational Collapse,'' \ajou Phys. Rev. &D14 (76) 2460.}
\lref\Mann{R.B. Mann, ``Lower dimensional gravity,'' Waterloo preprint
WATPHYS-TH-91-07, to appear in Proc. of 4th Canadian Conf. on General
Relativity and Relativistic Astrophysics, Winnipeg, Canada.}
\lref\tomb{E. Tomboulis, ``1/N expansion and renormalization in quantum
gravity,''  \ajou Phys. Lett. &B70 (77) 361.}
\lref\Jack{R. Jackiw, ``Liouville field theory: a two-dimensional model for
gravity?,'' in {\sl Quantum Theory of Gravity}, S. Christensen, ed.
(Hilger, Bristol U.K. 1984)
\semi A.H. Chamseddine, ``A solution to two dimensional quantum
gravity. Non-critical strings''\ajou Phys. Lett. &B256 (91) 379.}
\lref\CF{S. M. Christensen and S. A. Fulling, ``Trace anomalies and the
Hawking effect,''\ajou Phys. Rev. &D15 (77) 2088.}
\lref\AW{J. J. Atick and E. Witten, ``The Hagedorn transition and the number
of degrees of freedom of string theory'',\ajou Nucl. Phys. & B310 (88) 291;
W. Fischler and J. Polchinski, unpublished.}
\lref\BSW{M. Bowick, L. Smolin and R. Wijewardhana, Phys. Rev. Lett. ???}
\lref\PSSTW{J. Preskill, ``Quantum hair'', CALT-68-1671 (1990);
J. Preskill, P. Schwarz, A. Shapere, S. Trivedi and
F. Wilczek, ``Limitations on the statistical description of black holes,''
IAS preprint IASSSNS-HEP-91/34.}
\lref\DLP{L.J. Dixon, J. Lykken, and M.E. Peskin, ``N=2 superconformal
symmetry and SO(2,1) current algebra,''\ajou Nucl. Phys. & B325 (89) 329.}
\lref\GiMae{G.W. Gibbons, ``Antigravitating black hole solitons with scalar
hair in N=4 supergravity,''\ajou Nucl. Phys. &B207 (82) 337\semi
G.W. Gibbons and K. Maeda, ``Black holes and membranes in
higher-dimensional theories with dilaton fields,''\ajou Nucl. Phys. &B298
(88) 741}
\lref\BarsBH{I. Bars, ``String propagation on black holes,'' USC preprint
USC-91/HEP-B3; ``Curved space-time strings and black holes,'' USC preprint
USC-91/HEP-B4.}
\lref\ILS{N. Ishibashi, M. Li, and A.R. Steif, ``Two dimensional charged
black holes in string theory,'' UCSB preprint UCSB-91-28.}
\lref\EFR{S. Elitzur, A. Forge, and E. Rabinovici, ``Some global aspects of
string compactifications,'' Hebrew University preprint RI-143-90.}
\lref\RSS{M. Ro\v cek, K. Schoutens, and A.
Sevrin, ``Off-shell WZW models in extended superspace,'' IAS preprint
IASSNS-HEP-91-14.}
\lref\Witt{E. Witten, ``On string theory and black holes,''\ajou Phys. Rev.
&D44 (91) 314.}
\lref\CHS{C. Callan, J. Harvey and A. Strominger, ``Worldsheet
approach to
heterotic solitons and instantons,''\ajou Nucl. Phys. &B359 (91) 611.}
\lref\HoSt{D. Garfinkle, G. Horowitz, and A. Strominger, ``Charged black holes
in string theory,''\ajou Phys. Rev. &D43 (91) 3140\semi G. Horowitz and A.
Strominger, `` Black strings and 
$p$-branes,''\ajou Nucl. Phys. &B360 (91) 197.}
\lref\Hawk{S. W. Hawking, ``Particle creation by black holes,''
\ajou Comm. Math. Phys. &43 (75) 199.}
\lref\Waldrev{ For a recent review, see
R. Wald's 1991 Erice lectures , ``Black Hole Thermodynamics'', University of
Chicago preprint (1991). }
\lref\worm{S. Coleman, ``Black holes as red herrings: Topological
fluctuations and the loss of quantum coherence,''\ajou Nucl. Phys. &B307
(88) 867, S. B. Giddings and A. Strominger, ``Loss of incoherence and
determination of coupling constants in quantum gravity,''\ajou Nucl. Phys.
&B307 (88) 854.}
\lref\BaNe{I. Bars and D. Nemeschansky, ``String propagation in backgrounds
with curved space-time,''\ajou Nucl. Phys. &B348 (91) 89.}
\lref\Bars{I. Bars, ``Heterotic superstring vacua in four dimensions based
on non-compact affine current algebras,''\ajou Nucl. Phys. &B334 (90) 125.}
\lref\MSW{G. Mandal, A Sengupta, and S. Wadia, ``Classical solutions of 2d
string theory,'' IAS preprint IASSNS-HEP-91/10.}
\lref\MaSh{E.J. Martinec and S.L. Shatashvili, ``Black hole physics and
Liouville theory,'' Chicago preprint EFI-91-22.}
\lref\DVV{R. Dijkgraaf, H. Verlinde, and E. Verlinde, ``String propagation
in a
black hole geometry,'' Princeton/IAS preprint PUPT-1252=IASSNS-HEP-91/22.}
\lref\GiSt{S.B. Giddings and A. Strominger, ``Exact black fivebranes in
critical superstring theory,''\ajou Phys. Rev. Lett. &67 (91) 2930.}
\lref\HoHo{J.H. Horne and G.T. Horowitz, ``Exact black string solutions in
three dimensions,'' UCSB preprint UCSBTH-91-39.}
\Title{\vbox{\baselineskip12pt
\hbox{UCSB-TH-91-54}\hbox{EFI-91-67}\hbox{PUPT-1294}
\hbox{hepth@xxx--9111056}}}
{Evanescent Black Holes}
\baselineskip=12pt
\bigskip
\centerline{\ticp Curtis G. Callan, Jr.,$^*$ Steven B.
Giddings,$^{\dagger\ddagger}$}
\medskip
\medskip
\centerline{\ticp Jeffrey A. Harvey,$^\#$ and Andrew
Strominger$^{\dagger\natural}$}
\medskip

\bigskip
\centerline{\bf Abstract}
A renormalizable theory of quantum gravity coupled to a dilaton and
conformal matter in two space-time dimensions is analyzed. The theory is
shown to be exactly solvable classically.  Included among the exact
classical solutions are configurations describing the formation of a black
hole by collapsing matter.  The problem of Hawking radiation and
backreaction of the metric is analyzed to leading order in a $1/N$
expansion, where $N$ is the number of matter fields. The results suggest
that the collapsing matter radiates away all of its energy before an event
horizon has a chance to form, and black holes thereby disappear from the
quantum mechanical spectrum. It is argued that the matter asymptotically
approaches a zero-energy ``bound state'' which can carry global quantum
numbers and that a unitary $S$-matrix including such states should exist.
\bigskip
\bigskip
\medskip
{\sl $^*$Department of Physics, Princeton University,} {\sl Princeton, NJ
08544}

{\it Internet: cgc@pupphy.princeton.edu}

{\sl$^\dagger$Department of
Physics, University of California,} {\sl Santa Barbara, CA 93106}

{\it
$^\ddagger$Internet: giddings@denali.physics.ucsb.edu}

{\it
$^\natural$Bitnet: andy@voodoo}

{$\#$\sl Enrico Fermi Institute, University of Chicago, 5640 Ellis Avenue,

Chicago, IL 60637 }
{\it Internet: harvey@curie.uchicago.edu}

\Date{11/91}

Following his ground-breaking work \Hawk\ on black hole evaporation,
Hawking \unpred\ argued that the process of formation and subsequent
evaporation of a black hole is not governed by the usual laws of quantum
mechanics: rather, pure states evolve into mixed states \Waldrev.  This
conjecture is hard to check in detail because of the many degrees of
freedom and inherent complexity of the process in four spacetime
dimensions. It would be useful to have a toy model in which greater
analytic control is possible.

In this paper we investigate such a model. It is a consistent,
renormalizable theory of quantum gravity in two spacetime dimensions
coupled to conformal matter. It contains black hole solutions as well as
Hawking radiation, and is exactly soluble at the classical level.  As we
shall see, the theory is just complicated enough to enable one to ask the
interesting questions concerning black hole evaporation, yet simple enough
to obtain some answers.

We begin with the action in two spacetime dimensions
\eqn\one
{S= { 1 \over 2\pi}\int d^2 x\sqrt{-g}\left[e^{-2\phi}(R+4(\nabla\phi)^2
+4\lambda^2)
-\half(\nabla f)^2\right]}
where $g$, $\phi$ and $f$ are the metric, dilaton, and matter fields,
respectively, and $\lambda^2$ is a cosmological constant. This action
arises as the effective action describing the radial modes of extremal
dilatonic
black holes in four or higher dimensions\refs{\GiMae,\HoSt,\GiSt}\foot{In
the context of superstrings $f$ does not have the usual dilaton coupling
because it arises from a Ramond-Ramond field.};
it is also closely related to
the spacetime action for $c=1$ noncritical strings.  However, these
connections need not concern us here;
the theory defined by the action \one\ is of interest in its own right as a
renormalizable theory of two dimensional ``dilaton gravity'' coupled to
matter.

The quantization of related theories of $2D$ gravity has been considered in
\Jack. Gravitational collapse in related theories has been studied in \Mann.
The black hole solution of \one\ in the absence of matter has
appeared previously \Witt\ as a low-energy approximation to an exact
solution of string theory.

The classical theory described by \one\
is most easily analyzed in conformal gauge
\eqn\two
{\eqalign{g_{+-} &=-\half e^{2\rho}\cr
          g_{--} &= g_{++} = 0\cr}}
where $x^{\pm} =  (x^0\pm x^1)$. The metric equations of motion
then reduce to
\eqn\three
{\eqalign{T_{++} &=  e^{-2\phi}\left(  4\partial_+\rho
\partial_+\phi-2\partial^2_+ \phi\right) +\half \partial_+ f\partial_+f=0\cr
          T_{--} &=
 e^{-2\phi}\left(4\partial_-\rho\partial_-\phi-2\partial^2_-\phi\right)
 +\half\partial_-f\partial_-f=0\cr
          T_{+-} &=  e^{-2\phi}\left(2\partial_+\partial_-\phi -4
\partial_+\phi\partial_-\phi -\lambda^2e^{2\rho}\right)=0\ .\cr}}
The dilaton and matter equations are
\eqn\four{-4\partial_+\partial_-\phi + 4\partial_+\phi\partial_-\phi +
2\partial_+\partial_-\rho + \lambda^2 e^{2\rho}=0}
\eqn\five{\partial_+\partial_-f=0\ .}

The general solution of the dilaton, matter, and $T_{+-}$ equations
(which do not involve $f$) may be expressed in terms of two free fields
\eqn\six{\eqalign{w &= w_+(x^+) + w_-(x^-)\cr
         u&=u_+(x^+) +u_-(x^-)\cr}}
as
\eqn\seven{\eqalign{e^{-2\phi} &= u - h_+ h_-\cr
         e^{-2\rho} &= e^{-w}(u - h_+ h_-)\cr}}
where
\eqn\eight{h_\pm (x^\pm) = \lambda \int\limits^{x^\pm}e^{w_\pm}\ .}
The matter equation of course implies
\eqn\nine{f=f_+ (x^+) + f_-(x^-)\ .}
The remaining constraint equations $T_{++}=T_{--}=0$
may then be solved for $u$ in terms
of $f_\pm$ and $w_\pm$. The general solution is
\eqn\ten
{u_\pm = {M\over2\lambda }  -\half \int\ e^{w_\pm}\int\ e^{-w_\pm} \partial_\pm
f\partial_\pm f  .}
where $M$ is an integration constant.

We now consider solutions with $f=0$, which implies that one can set
$u=M/\lambda$. Conformal
gauge leaves unfixed the conformal subgroup of diffeomorphisms.  This
gauge freedom can be fixed (on-shell) by setting $w=0$.  The general $f=0$
solution
is then
\eqn\eleven{\eqalign{e^{-2\phi} &={M\over\lambda}-\lambda^2 x^+ x^-\cr
                     &= e^{-2\rho}\cr}}
up to constant translations of $x$. It is readily
seen \GiSt\ that for $M\not=0$, this corresponds to the $r-t$ plane of
the of the
higher dimensional black holes of \refs{\GiMae,\HoSt} near the extremal
limit, or to the two dimensional
black hole solution of \Witt, with $M$ the black hole mass.
It is not immediately apparent that the parameter $M$ corresponds to the
black hole mass.  This can be verified by a calculation of the ADM
mass for this configuration as described in \Witt\ or by a
calculation of the Bondi mass as is done later in this paper.
For $M=0$ one can introduce coordinates in which the metric is flat and
the dilaton field $\phi$ is linear in the spatial coordinate.
This ``linear dilaton'' vacuum has appeared in previous studies
of lower-dimensional string theories and also corresponds to extremal
higher-dimensional black holes.

{}From the  above we may expect that any matter perturbation of the linear
dilaton vacuum will result in the formation of a black hole. To see that
this is indeed the case consider the example of
an $f$ shock-wave traveling in the $x^-$ direction with magnitude $a$
described by the stress tensor
\eqn\twelve
{\half \partial_+ f\partial_+ f= a\delta (x^+ - x^+_0)\ .}
One then finds in the gauge $w=0$ that
\eqn\thirteen{ e^{-2\rho} = e^{-2\phi} =-a(x^+ - x^+_0)
\Theta(x^+ - x^+_0) -\lambda^2 x^+ x^- . }
For $x^+<x^+_0$, this is  simply the linear dilaton vacuum while for
$x^+ > x^+_0$ it is identical to a black hole of mass $ax^+_0 \lambda$ after
shifting $x^-$ by $a/\lambda^2$.
The two solutions are
are joined along the $f$-wave.  The Penrose diagram for this
spacetime is depicted in Figure 1.
\goodbreak
\topinsert
\centerline{\psfig{figure=ebh.fig1}}
\endinsert

The fact that any $f$-wave, no matter how weak, produces
a black hole of course implies that weak field perturbation theory
breaks down. The reason for this is simple. From \one\ it is
evident that the weak field expansion parameter is proportional to
$e^{\phi}$. Equation \thirteen\ shows that this parameter becomes arbitrarily
large close to ${\cal I}_L^\pm$ or to the singularity and that
the weak field expansion diverges in this region.

This has a higher dimensional interpretation
as follows \HoSt. When \one\ is taken
as an effective field theory for higher-dimensional dilatonic black
holes, the $2D$ linear dilaton vacuum corresponds to the infinite throat in
the extremal black hole solutions.
The center of the black hole is at $x^+x^-=0$.
An arbitrarily small infalling matter wave then produces a non-extremal
black hole with an event horizon and a singularity.

So far the discussion has been purely classical.  As a first step
towards including quantum effects, we now compute the Hawking radiation
in the fixed background geometry \eleven.  This can be computed exactly for
the collapsing $f$-wave because of the elegant relation \CF\ between
Hawking radiation and the trace anomaly for $2D$ conformal
matter coupled to gravity.

The calculation and its physical interpretation is clearest in coordinates
where the metric is asymptotically constant on ${\cal I}_R^{\pm}$.
We thus set
\eqn\eighteen
{\eqalign{e^{\lambda\sigma^+}& =\lambda x^+ \cr
          e^{-\lambda\sigma^-} &= -\lambda x^- - {a\over\lambda}.\cr}}
This preserves the conformal gauge \two\ and gives for the new
metric
\eqn\newmet{-2 g_{+-} = e^{2 \rho} = \cases{[1+{a \over \lambda} e^{\lambda
\sigma^-}]^{-1},
             & if $\sigma^+ < \sigma_0^+$; \cr
            [1+ {a \over \lambda} e^{\lambda (\sigma^- -\sigma^+ +\sigma_0^+)}
]^{-1}
             & if $\sigma^+ > \sigma_0^+$ \cr}}
with $\lambda x_0^+ = e^{\lambda \sigma_0^+}$.
By the standard one-loop anomaly argument,
the trace $T^f_{+-}$ of the stress tensor is
proportional to the curvature scalar which is, in these coordinates, just the
laplacian of $\rho$. The result is\foot{It is assumed that the functional
measure for the matter fields is defined with the metric $g$. One could
imagine using instead
the (flat) metric $e^{-2\phi}g$, in which case there would
be no Hawking radiation.}
\eqn\fourteen
{\vev{T^f_{+-}} =-{1 \over 12} \partial_+\partial_-\rho~~.}
One can then integrate the equations of conservation of $T^f$
to infer the following one-loop contributions to $T_{++}^f$ and $T_{--}^f$:
\eqn\fifteen
{\eqalign{\vev{T^f_{++}}& = -{1 \over 12} \left(
 \partial_+\rho \partial_+\rho - \partial^2_+\rho +
t_+(\sigma^+)\right)\ ,\cr
          \vev{T^f_{--}}& =-{1 \over 12} \left(
  \partial_-\rho\partial_-\rho - \partial^2_-\rho +
t_-(\sigma^-)\right)\ .\cr}}
The functions of integration $t_\pm$ must be fixed by boundary conditions.
For the collapsing
$f$-wave, $T^f$ should vanish identically in the linear dilaton region, and
there should be no incoming radiation along ${\cal I}^-_R$ except for the
classical $f$-wave at $\sigma^+_0$.  Using the formula for $\rho$, this
implies
\eqn\sixteen
{t_+ = 0, \qquad t_- = {-\lambda^2 \over 4} [1- (1+a e^{\lambda
\sigma^-}/\lambda )^{-2} ]. }
The stress tensor is now completely determined, and one can read off
its values on ${\cal I}^+_R$ by taking the limit $\sigma^+\to \infty$:
\eqn\nineteen
{\eqalign{ \vev{T^f_{++}} &\to 0 \qquad\vev{T^f_{+-} }\to 0\cr
          \vev{T^f_{--}} &\to {\lambda^2 \over48} \left[
1-{1\over\left(1+a e^{\lambda\sigma^-}/\lambda\right)^2}\right]~~.}}
The limiting value of $T^f_{--}$ is the flux of f-particle energy across
${\cal I}^+_R$. In the far past of ${\cal I}^+_R$ $(\sigma^- \to-\infty)$ this
flux vanishes exponentially while, as the horizon is approached, it
approaches the constant value $\lambda^2/48$.
This is nothing but Hawking radiation. The surprising result
that the Hawking radiation rate is asymptotically
independent of mass has been found in
other studies of two-dimensional gravity.

The total energy lost by the collapsing $f$-wave at some value of
retarded time $\sigma^-$ can be estimated by integrating the outgoing flux
along ${\cal I}^+_R$ up to $\sigma^-$. If the total radiated flux is computed
by
integrating along all of ${\cal I}^+_R$, an infinite answer is obtained,
because
the outgoing flux approaches a steady state at late retarded times.  This
is obviously nonsense --- the black hole cannot lose more mass than it
possesses.  This nonsensical answer is, of course, a result of the fact that
we have neglected the backreaction of the radiation on the collapsing
$f$-wave.  As a first step toward analyzing the backreaction, it is
useful to estimate, to leading order in the
mass, the retarded time at which the
integrated energy of the Hawking radiation on ${\cal I}^+_R$ equals the initial
mass $ax_0^+\lambda$ of the incoming $f$-wave.  This is given by
%
\eqn\nty{e^{-\lambda \sigma^-} = {e^{-\lambda \sigma_0^+} \over 24}. }
By this time, Hawking radiation has backscattered
all the energy of the incoming $f$-wave into outgoing flux on ${\cal I}^+_R$.

Unfortunately this picture cannot yet be taken seriously because
the turn-around point at which all the energy has backscattered has coordinates
$(\sigma_0^+ , \sigma_0^+ + (\log 24)/\lambda)$.
The value of the dilaton at this point is from \thirteen\ for small mass
\eqn\ntyone{e^{-2\phi}={1 \over 24}  }
independent of $\sigma^+_0$ or $a$.   As we have stated,
$e^\phi$ is the loop expansion parameter for dilaton
gravity.  Since this parameter  is not small at the turn-around point,
our one-loop calculation
of the Hawking flux breaks down before the $f$-wave fully
backscatters.

The situation can be remedied by proliferating the number of matter fields.
This introduces a new small expansion parameter into the theory:
$1/N$, where $N$ is the (large) number of matter fields \tomb.
For $N$ matter fields the Hawking flux is $N$ times as
great and one finds that the $f$-wave has completely backscattered by
$ (\sigma_0^+ , \sigma_0^+ + (\log 24/N)/\lambda )$.
For large $N$, the value of the dilaton at this point is
\eqn\ntytwo{e^{-2\phi}={N \over 24}}
which indeed corresponds to weak coupling. This suggests that, for large N,
the essential physics of Hawking radiation backreaction takes place in a
weak coupling regime and should be amenable to a semiclassical treatment.
In what follows, we will present some proposals for the development of
such a fully consistent treatment of the scattering problem, along with
some informed conjectures about the form of the solution.

In a systematic expansion in $\frac{1}{N}$, one must include the one-loop
matter-induced contribution to the gravitational effective action at the
same order as the classical action \one.  This incorporates both Hawking
radiation and backreaction.  Because of
the way the dilaton varies with position, there is a region in spacetime
where the $O(N)$ one-loop matter-induced gravitational action is of the
same order as the strictly classical part and the loop coupling constant
is nonetheless small. As described above, it is precisely in this region
where the essential backreaction physics will occur and a semiclassical
treatment of the proper action should give meaningful answers.
To leading order in $\frac{1}{N}$, and in conformal gauge,
the quantum effective action to be solved is
\eqn\ntythree{\eqalign{S_N &= {1 \over \pi}\int\ d^2\sigma\ \Bigl[e^{-2\phi}
(-2\partial_+ \partial_-\rho + 4\partial_+\phi\partial_-\phi
-\lambda^2 e^{2\rho})\cr
             & - \half\sum\limits^N_{i=1} \partial_+ f_i\partial_-
f_i + {N \over 12} \partial_+\rho\partial_-\rho\Bigr]~.\cr}}
The last term is the Liouville term induced by the $N$ matter
fields and the conformal gauge constraints (the $T_{\pm\pm}$ equations of
\three ) are modified by its presence in a way which will shortly
be made explicit\foot{In a systematic quantum treatment of this action
one will find, at subleading order in $1/N$, that the $N$ in \ntythree\
is shifted by the ghost and gravity measures in order to maintain a
net central charge $c=26$.}. We have also tuned the coefficient
of the possible Liouville cosmological constant (to be distinguished
from the classical ``dilatonic'' cosmological constant $\lambda$)
to zero. In a slight abuse of terminology, we nevertheless
refer to the dynamics governed by the last term in \ntythree\ as
Liouville gravity. Solving the quantum theory to leading order
in $1\over N$ is equivalent to solving the classical theory
described by $S_N$.

Unlike $S_0$, it does not appear possible to solve $S_N$ exactly,
though it may be possible to solve the equations numerically.
At present the best we can do is make
the following educated guess about the evolution of an incoming $f$-wave.
Consider a quantization of the system
defined on null surfaces $\Sigma(\sigma^-)$ of constant $\sigma^-$.  The
light-cone Hamiltonian $P_-$
evolves the system in the direction of increasing $\sigma^-$.  The charges
$P_\pm$ are not separately conserved because translation invariance is
spontaneously broken. The combination $H=P_++P_-$ generates an unbroken
symmetry and is conserved for spacelike surfaces. In fact there are in
general two conserved quantities, given by boundary terms at
the two spatial infinities. For the null surfaces $\Sigma$,
the eigenvalue $M(\sigma^-)$ of $H$ is given by a boundary term on ${\cal
I}^+_R$
(assuming the boundary term on ${\cal I}^-_L$ vanishes) and is called the
Bondi mass.  The Bondi mass is not conserved because radiation energy
can leak out onto ${\cal I}^+_R$.

Now consider an initial state at $\sigma^-=-\infty$ describing an incoming
$f_1$-wave as in \twelve, with the other $N-1$ $f$'s set to zero.
In addition it is useful to let
this wave be characterized by the non-anomalous,
left-moving global conserved charge $Q_{1L}=\int d \sigma^+ \partial_+f_1$.
Near $\sigma^-=-\infty$, $e^{-2\phi}$ is very
large  and the extra Liouville term may therefore be
neglected in the description of the incoming state on $\Sigma (-\infty)$, which
is essentially described by \thirteen.
As $\sigma^-$ increases away from ${\cal I}^-_R$, $M(\sigma^-)$ will
decrease.  From the point of view of the quantum effective action $S_N$,
this is not due to Hawking radiation, but is simply a consequence of the
extra Liouville term.  As $\sigma^-\to +\infty$, it is
plausible that $M(\sigma^-)$ decreases to zero.
However, the state on $\Sigma (\sigma^-)$ can not revert to the linear dilaton
vacuum on ${\cal I}^+_L$ because it carries the conserved charge $Q_{1L}$.

The picture can thus be summarized as  follows.  A state with non-zero
charge $Q_{1L}$ and
Bondi energy is incoming from ${\cal I}^-_R$.  As it evolves it loses its
energy, but retains its charge.  Asymptotically it approaches a
zero-energy state with charge $Q_{1L}$ on ${\cal I}^+_L$. This
is illustrated in Figure 2.
\goodbreak
\topinsert
\centerline{\psfig{figure=ebh.fig2}}
\endinsert

This picture can be corroborated by direct analysis of the Bondi energy
associated with data on a null surface $\Sigma$ corresponding to a charged
$f$-wave.  Such data must satisfy the null constraint equations:
\eqn\ntyfour
{\eqalign{0=T_{++}& = e^{-2\phi} (4\partial_+\phi\partial_+\rho -
2\partial^2_+\phi) + \half \partial_+ f\partial_+f\cr
                  & - {N \over 12}\left(\partial_+\rho\partial_+\rho -
\partial^2_+\rho + t_+(\sigma^+)\right)\cr
          0=T_{+-} & = e^{-2\phi}(2\partial_+\partial_-\phi - 4
\partial_+\phi\partial_-\phi - \lambda^2e^{2\rho})\cr
                   &- {N \over 12} \partial_+\partial_-\rho\ .\cr}}
The extra function $t_+$ appearing in $T_{++}$ is in agreement with
\fifteen\ and is a consequence of the
anomalous transformation law for $T_{++}$. $t_+$ is coordinate dependent
and must be fixed by boundary conditions, as in \sixteen.
The linear dilaton configuration remains as the vacuum solution of the
full leading $N$ theory:
\eqn\ntyfive{\eqalign
{\rho&=0\cr
 f_i&=0\cr
 \phi&=-\frac{\lambda}{2} (\sigma^+-\sigma^-).\cr}}
The Bondi energy may then be defined for
configurations which approach \ntyfive\ on ${\cal I}^+_R$ (i.e. the
configuration must not only be asymptotically flat, but presented in
an asymptotically Minkowskian coordinate system).  It is
given by the surface term which must be added to the integral of
$T_{++} +T_{+-}$ over $\Sigma$ to obtain the generator of time
translations. This canonical procedure yields
\eqn\ntysix{
\eqalign{M(\sigma^-)&=2 e^{\lambda(\sigma^+ - \sigma^-)} (\lambda \delta \rho +
               \partial_+ \delta \phi - \partial_- \delta \phi)\cr
          &+ {N \over 12} (\partial_- \delta \rho-\partial_+ \delta \rho)\cr}}
where $\delta \rho$ and $\delta \phi$, are the asymptotically vanishing
deviations of $\rho$ and $\phi$ from \ntyfive,
and the right hand side is to be evaluated on ${\cal
I}^+_R$. The first ``dilaton'' term was obtained in reference \Witt.
The term proportional to $N$, arising from matter quantum effects,
actually vanishes due to the boundary conditions \ntyfive.
A modified formula is required in coordinate systems (such as in \eleven)
for which the fields do not asymptotically approach \ntyfive.

Let us first consider the energy, evaluated on a surface $\Sigma$,
of a small amplitude $f_1$-wave packet
localized in the `dilaton region' where $e^{-2\phi}$ is very large,
i.e. at very large
$\sigma^+-\sigma^-$. Then the Liouville terms proportional to $N$ may be
neglected in solving the constraints.  $M$ will be given as before by
the integrated value of $\half \partial_+ f_1\partial _+ f_1$ times the $x^+$
coordinate of the center of the wave packet.

Now, however, consider the case where the $f_1$ wave-packet is localized
on $\Sigma$ in the `Liouville region' where $e^{-2\phi}$ is very small.
The dilaton gravity term
is then very small, and the action governing $\rho$ and $f$ \foot{The
dynamics of $\phi$ are roughly governed by the free field $\psi=e^{-\phi}$.
However it is not clear what range should be taken for $\psi$.}
reduces to Liouville gravity coupled
to conformal matter:
\eqn\ntyseven
{S_N({\rm large}\ \phi) ={1 \over \pi}\int
 d^2\sigma\left( {N \over 12}\partial_+\rho\partial_-\rho - \half
\sum\limits^N_{i=1}\partial_+f_i\partial_-f_i\right)}
with constraints
\eqn\ntyeight{\eqalign
{0=T_{++}&=\half\sum\limits^N_{i=1}\partial_+f_i\partial_+f_i\cr
         &- {N \over 12}
\left(\partial_+\rho\partial_+\rho - \partial^2_+\rho +t_+(\sigma^+)\right)\cr
 0=T_{+-}&=-{N \over 12} \partial_+\partial_-\rho\ .}}
The $T_{+-}$ constraint implies that the spacetime is in fact flat.  The
Bondi energy of \ntysix\ reduces to its Liouville piece
\eqn\ntynine{M( \sigma^- )={N \over 12}
(\partial_- \delta \rho-\partial_+ \delta \rho)~.}
Since there is no invariant one can associate to a flat
metric one would expect this expression to vanish. That it does can
be seen from direct evaluation of \ntynine:
if $\rho$ approaches zero on ${\cal I}^+_R$ as
required by the boundary conditions \ntyfive, the derivatives of $\rho$ and
consequently $M$ must also vanish. Thus all asymptotically flat states
of Liouville gravity plus matter have zero energy.

We now have a plausible global picture of the scattering process.  The linear
dilaton vacuum is divided into two regions characterized by $e^{-2\phi}$
large or small compared to ${N \over 12}$.  This dividing line is timelike.
For $e^{-2\phi} \gg {N \over 12}$, the dynamics are essentially that of
classical dilaton gravity coupled to matter.
For $e^{-2\phi}\ll {N \over 12}$, one has Liouville
gravity coupled to matter.  An incoming $f_1$ wave-packet on ${\cal I}^-_R$
begins in the dilaton gravity region where it has non-zero Bondi energy.
However, it eventually crosses into the Liouville region, where all
excitations have zero energy. By energy conservation, all of the initial
energy of the wavepacket must have radiated away to ${\cal I}^+_R$.
There is no indication of an event horizon or singularity\foot{In
two dimensions, unlike in higher dimensions,
we know of no local notion of an apparent horizon. Global event horizons
exist as usual when the spacetime is singular or otherwise
incomplete.}: in the region where
the singularity occurs in the classical solution, the quantum dynamics are
governed by Liouville gravity (with no cosmological constant) in which
the curvature is required to vanish. One expects,
therefore, a unitary $S$-matrix evolving from ${\cal I}^- $ to
${\cal I}^+ $. One would hope to extract information about this
$S$-matrix from a semiclassical treatment of the
large-N action \ntythree.

While we find this picture compelling, we
emphasize that it must at present be regarded as speculative.
We have $not$ shown that an incoming $f$-wave does not in fact produce
a singularity, or even that the large $N$ equations of motion
give a well-defined evolution. One might
try to substantiate our speculations by doing
weak field perturbation theory in the amplitude of the $f$-wave.
However preliminary calculations indicate that weak field perturbation
theory breaks down near the boundary of the dilaton and Liouville regions:
the second-order perturbation is divergent.  Thus in order to
settle the question a non-perturbative analysis of the large $N$
theory \ntythree\ is probably needed\foot{Of course a singularity at large $N$
does not imply a singularity of the full quantum theory since the
$1/N$ expansion breaks down as soon as fields grow to order $N$.}.

In conclusion we have analyzed the process of black hole formation
and evaporation, including backreaction, in the $1/N$ expansion of a
two-dimensional model.  A set of equations describing the process were
found, but have so far not been solved. A qualitative analysis
suggests that in this model would-be black holes in fact evaporate
before an event horizon or singularity has a chance to form. Thus
there is no indication that pure states evolve
into mixed states. The implications of our results for four-dimensional
black holes remain to be explored.

\bigskip\centerline{\bf Acknowledgements}\nobreak
We are grateful to S. Hawking, G. Horowitz, J. Preskill and R. Wald for useful
discussions. After completion of this work we learned that some aspects of
this theory have been considered by E. and H. Verlinde.

This work was supported in part by DOE grant DE-AC02-84-1553,
DOE grant DE-AT03-76ER70023, and NSF grant PHY90-00386.
S.B.G. also acknowledges the support of NSF PYI grant PHY-9157463
and J.A.H. acknowledges the support of NSF PYI grant PHY-9196117.

\listrefs
\end